\documentclass[a4paper, amsfonts, amssymb, amsmath, reprint, showkeys, nofootinbib, twoside]{revtex4-1}
\usepackage[english]{babel}
\usepackage[utf8]{inputenc}
\usepackage{svg}
\usepackage{xcolor}
\usepackage[colorinlistoftodos, color=green!40, prependcaption]{todonotes}
\usepackage{amsthm}
\usepackage{mathtools}
\usepackage{physics}
\usepackage{xcolor}
\usepackage{graphicx}
\usepackage[left=23mm,right=13mm,top=35mm,columnsep=15pt]{geometry} 
\usepackage{adjustbox}
\usepackage{placeins}
\usepackage[T1]{fontenc}
\usepackage{lipsum}
\usepackage{csquotes}
\usepackage[pdftex, pdftitle={Article}, pdfauthor={Author}]{hyperref} 

\usepackage{comment}

\begin{document}
\title{Tuning the Charge Transfer of Transition Metal Dichalcogenides via Misfit Layer Compounds}

\author{Hugo Le Du$^{1, \ast,\dagger}$, Ludovica Zullo$^{2, \ast, \dagger}$, Justine Cordiez$^{3, \ast}$, Robin Salvatore$^{1}$, Giovanni Marini$^{4}$, Marie Hervé$^{1}$, Debora Pierucci$^{1}$, Shunsuke Sasaki$^{3}$, Florent Pawula$^{3}$, Etienne Janod$^{3}$, Chiara Bigi$^5$, Marta Zonno$^5$, François Bertran$^5$, Thomas Jaouen$^6$, Patrick Le Fèvre$^6$, Matteo Calandra$^{4}$, Laurent Cario$^{3}$, Tristan Cren$^{1}$ and  Marie D'Angelo$^{1, \dagger}$}
\affiliation{\vspace{0.3cm}
    \begin{tabular}{c}
        $^1$Institut des Nanosciences de Paris, Paris, France \\
        $^2$Institut f{\"u}r Theoretische Physik und Astrophysik and W{\"u}rzburg-Dresden Cluster of Excellence ctd.qmat, \\ Universität W{\"u}rzburg, 97074 W{\"u}rzburg, Germany \\
        $^3$Institut des Matériaux de Nantes Jean Rouxel, Nantes, France \\
        $^4$University of Trento, Italy \\
        $^5$Synchrotron SOLEIL, L’Orme des Merisiers, Saint-Aubin, France \\
        $^6$Institut de Physique de Rennes, France \\
        \noalign{\smallskip}
        \small $^\ast$ These authors contributed equally to this work \\
        \small $^\dagger$ Corresponding authors: hugo.ledu@insp.jussieu.fr, ludovica.zullo@uni-wuerzburg.de, dangelo@insp.jussieu.fr
    \end{tabular}
}

\begin{abstract}
Misfit layer compounds (MLCs) are a versatile platform for exploring the electronic phase diagram of two-dimensional (2D) materials beyond the limits of conventional gating techniques. This work demonstrates the precise tunability of electron doping in NbSe$_2$ monolayers through chemical alloying within the rocksalt layer of $(\text{La}_x\text{Pb}_{1-x}\text{Se})_{1.14}(\text{NbSe}_2)_2$ heterostructures. By combining first-principles density functional theory (DFT) calculations with angle-resolved photoemission spectroscopy (ARPES), we prove that the rocksalt unit acts as an universal electron donor. 
We show that varying the La/Pb ratio results in a rigid Fermi level shift, still preserving the NbSe$_2$ electronic structure. 
Crucially, photon-energy-dependent ARPES confirms that the NbSe$_2$ layers nearly maintain their intrinsic 2D character and orbital identity within the three-dimensional misfit. This study establishes MLCs as a reliable platform for engineering emergent states in 2D transition metal dichalcogenides through precise stoichiometric control.
\end{abstract}

\maketitle

\section{Introduction} \label{sec:Introduction}

The family of transition metal dichalcogenides (TMDs), characterized by the general formula MX$_2$ (where M denotes a transition metal and X a chalcogen), has emerged as a versatile platform for investigating correlated electronic phases in two dimensions, including charge density waves (CDWs) and superconductivity \cite{wang_electronics_2012,coleman_scanning_1988, manzeli2017, hwang_charge_2024, dobrovolskiy_roadmap_2026}. 

Owing to the strong interplay between electronic structure and many-body interactions, the physical properties of TMDs are remarkably sensitive to carrier density \cite{xu2021, bin2021, jung2023, orsted2025, lu2018}. Yet, conventional electrostatic tuning approaches, such as field-effect transistor (FET) gating, remain intrinsically constrained by a maximum achievable charge accumulation on the order of $1 \times 10^{14} \,~ \text{cm}^{-2}$\cite{ye_superconducting_2012}, thereby limiting access to broader regions of the phase diagram.

\begin{figure}
    \includegraphics[scale=0.35]{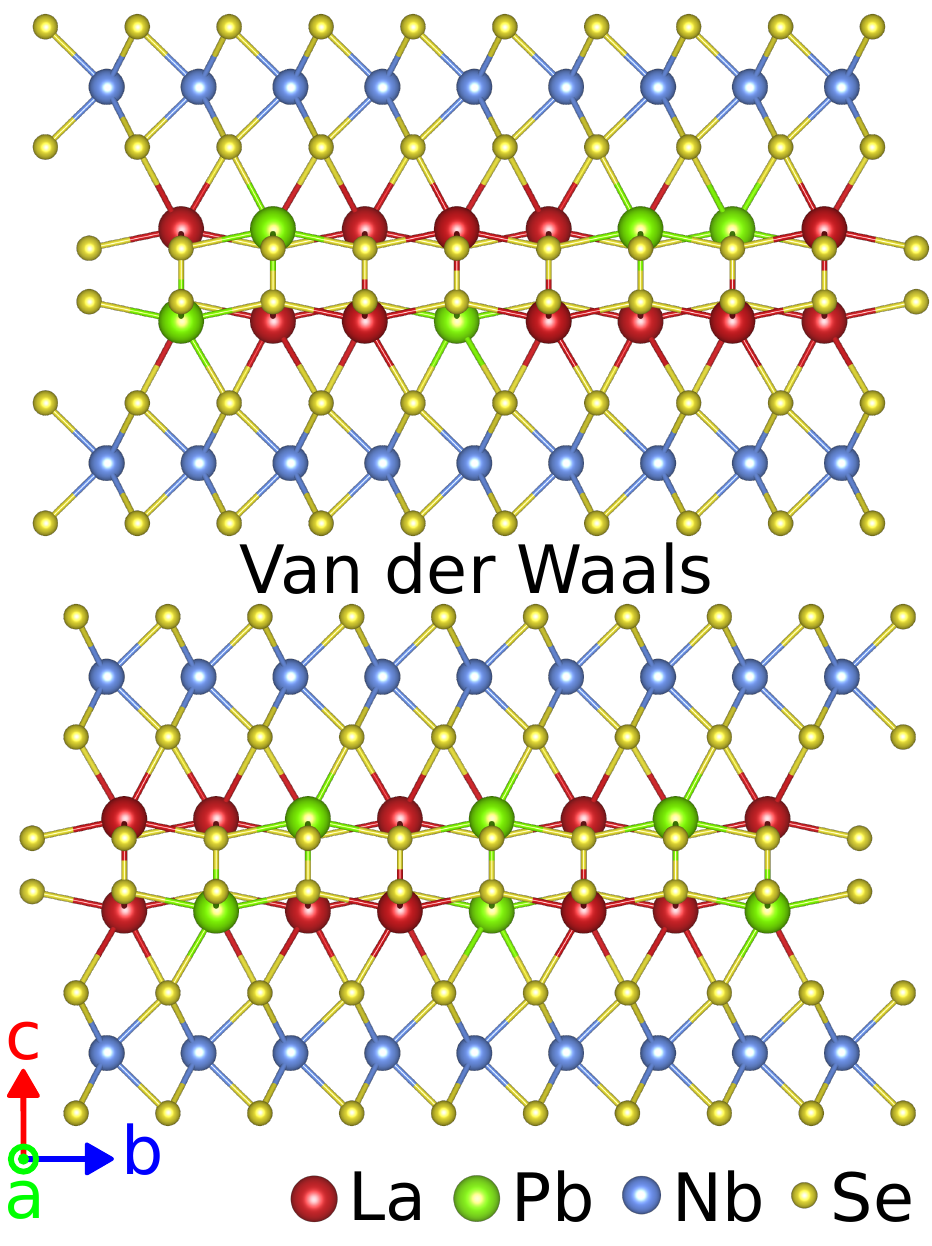} 
    \caption{Crystallographic structure of the misfit compound $(\text{La}_{x}\text{Pb}_{1-x}\text{Se})_{1.14}(\text{NbSe}_2)_2$. Strong iono-covalent bonds link the RS layers $\text{La}_x\text{Pb}_{1-x}\text{Se}$ and TMD layers $\text{NbSe}_2$ whereas the TMD layers are held together by weak van der Waals forces.}
    \label{fig:structure}
\end{figure}

To explore the limit of heavy electron-doping in monolayer TMDs beyond these constraints, misfit layer compounds (MLCs) have emerged as a powerful  platform \cite{leriche_misfit_2021}. 
These heterostructures are formed by the alternating stacking of rocksalt (RS) chalcogenide layers and TMD units \cite{ng_misfit_2022, wiegers_misfit_1993, Rouxel01041994}.
A sketch of a misfit layer compound structure is depicted in Fig. \ref{fig:structure}.
Weak Van der Waals forces control the cohesion between TMD layers, while strong iono-covalent bonding characterizes the interface between the RS and TMD layers. Therefore, cleaving a misfit between its two stacked TMD layers can yield clean surfaces terminated by a single TMD layer.

Among the possible TMD candidates to build a misfit heterostructure, NbSe$_2$ has emerged as an interesting platform to explore phenomena like charge density waves and superconductivity \cite{Revolinsky1965,Moncton1977a,Malliakas2013}. 
NbSe$_2$ typically crystallizes in a 2H stacking sequence that preserves global inversion symmetry. However, the reduction to a single 1H monolayer explicitly breaks this spatial inversion symmetry. 
Remarkably, in monolayer NbSe2, the absence of inversion symmetry, combined with strong spin–orbit coupling, generates an unconventional electronic structure characterized by out-of-plane spin-momentum locking \cite{xiao_coupled_2012, samuely_2023}, the so-called Ising spin-orbit coupling. This unique electronic landscape gives rise to Ising superconductivity, in which the in-plane upper critical magnetic field significantly exceeds the conventional Pauli paramagnetic limit \cite{xi_ising_2016, samuely_2021, engstrom_2025}. 
In this global context, extending the achievable doping level in TMDs is of crucial importance in order to explore further their fascinating properties.

Experimental investigations of the misfit $(\text{LaSe})_{1.14}(\text{NbSe}_2)_2$ have provided direct evidence of an extreme doping regime \cite{leriche_misfit_2021}. Scanning tunneling microscopy (STM) and angle-resolved photoemission spectroscopy (ARPES) measurements, supported by first-principles calculations, demonstrate that the NbSe$_2$ layers within the misfit behave as rigidly doped monolayers. In this compound, a charge transfer of approximately 0.55--0.6 electrons per Nb atom is achieved, corresponding to a Fermi level shift of $0.3$ eV, a regime several times larger than that accessible by conventional ionic liquid gating \cite{ye_superconducting_2012}.

However, the fundamental question remains as to how MLCs can be manipulated so as to precisely control the charge transfer. Recent theoretical advancements have further generalized the potential of MLCs, characterizing them as arrangements of ultratunable field-effect transistors \cite{zullo_misfit_2023}. Because TMDs generally possess higher work functions than rocksalt units, the RS layers act as universal donors while the TMDs serve as acceptors. Crucially, this charge transfer can be precisely controlled through chemical alloying within the rocksalt layer. For instance, theoretical calculations showed that by substituting La$^{3+}$ with Pb$^{2+}$ in systems like $(\text{La}_x\text{Pb}_{1-x}\text{Se})_{1.18}(\text{TiSe}_2)_2$ allows for the continuous tuning of the Fermi level \cite{zullo_misfit_2023}. This tunability not only enables systematic exploration and mapping of complex electronic phase diagrams, but also provides a route to the design of emergent properties, for example by inducing superconductivity via band structure engineering.

In this work, we confirm experimentally the predicted theoretical tunability. We present a complete study of the $(\text{La}_{x}\text{Pb}_{1-x}\text{Se})_{1.14}(\text{NbSe}_2)_2$ misfit series.
By means of first principles calculations we are able to accurately describe the electronic properties and charge transfer of this misfit series as a function of relative La-Pb concentration.
Motivated by our theoretical predictions, misfit crystals are synthesized via chemical vapor transport.
Finally, by using ARPES, we directly probe the evolution of the electronic band structure as a function of lanthanum concentration. Our results demonstrate that control of stoichiometry allows achieving intermediate doping levels between the heavily n-doped $(\text{LaSe})_{1.14}(\text{NbSe}_2)_2$ limit and isolated NbSe$_2$. Crucially, we confirm that the quasi-2D nature of the NbSe$_2$ layers is preserved throughout the substitution range, validating the field-effect transistors modeling of misfit layer compounds and establishing these materials as an ideal platform for exploring phase diagrams under precisely controlled doping.

\section{Results and Discussion} \label{sec:Results}

    \begin{figure*}
        \includegraphics[scale=0.35]{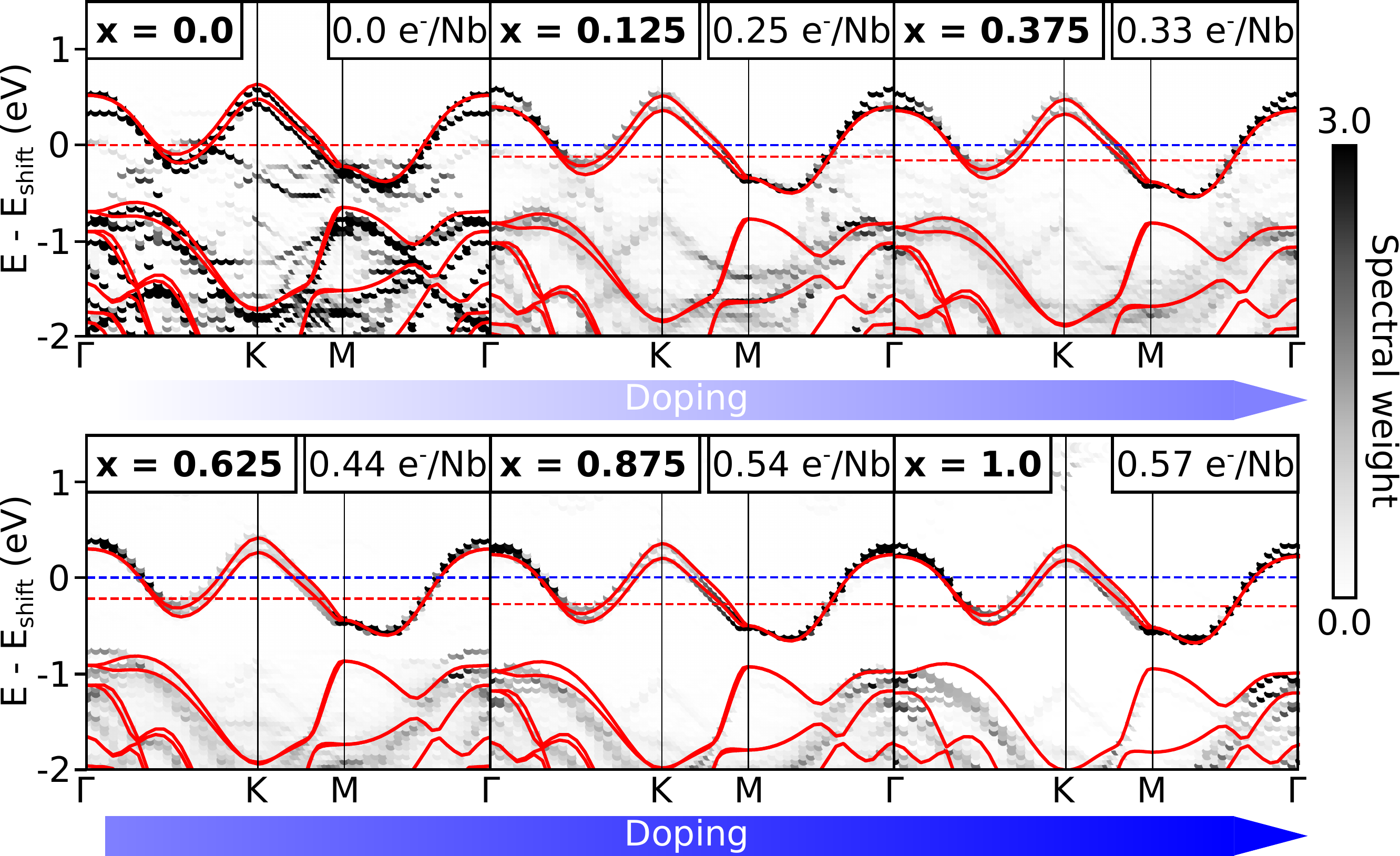} 
            \caption{Band unfolding onto the single layer NbSe$_2$ Brillouin zone of the misfit compound surfaces $(\text{La}_{x}\text{Pb}_{1-x}\text{Se})_{1.14}(\text{NbSe}_2)_2$ for $x = 1, 0.875, 0.625, 0.375,  0.125$ and $0$. The color scale follows the spectral weight of the band. The band structure of the isolated single layer $\text{NbSe}_2$ (red line) is superimposed and aligned to the Nb d-band in the misfit. The blue dashed line corresponds to the Fermi level $E_{F}$ of the misfit compound, while the red one to the Fermi level of the isolated $\text{NbSe}_2$ layer (in the first panel they are superimposed).}
        \label{fig:band}
    \end{figure*}

    \begin{figure*}
        \includegraphics[scale=0.35]{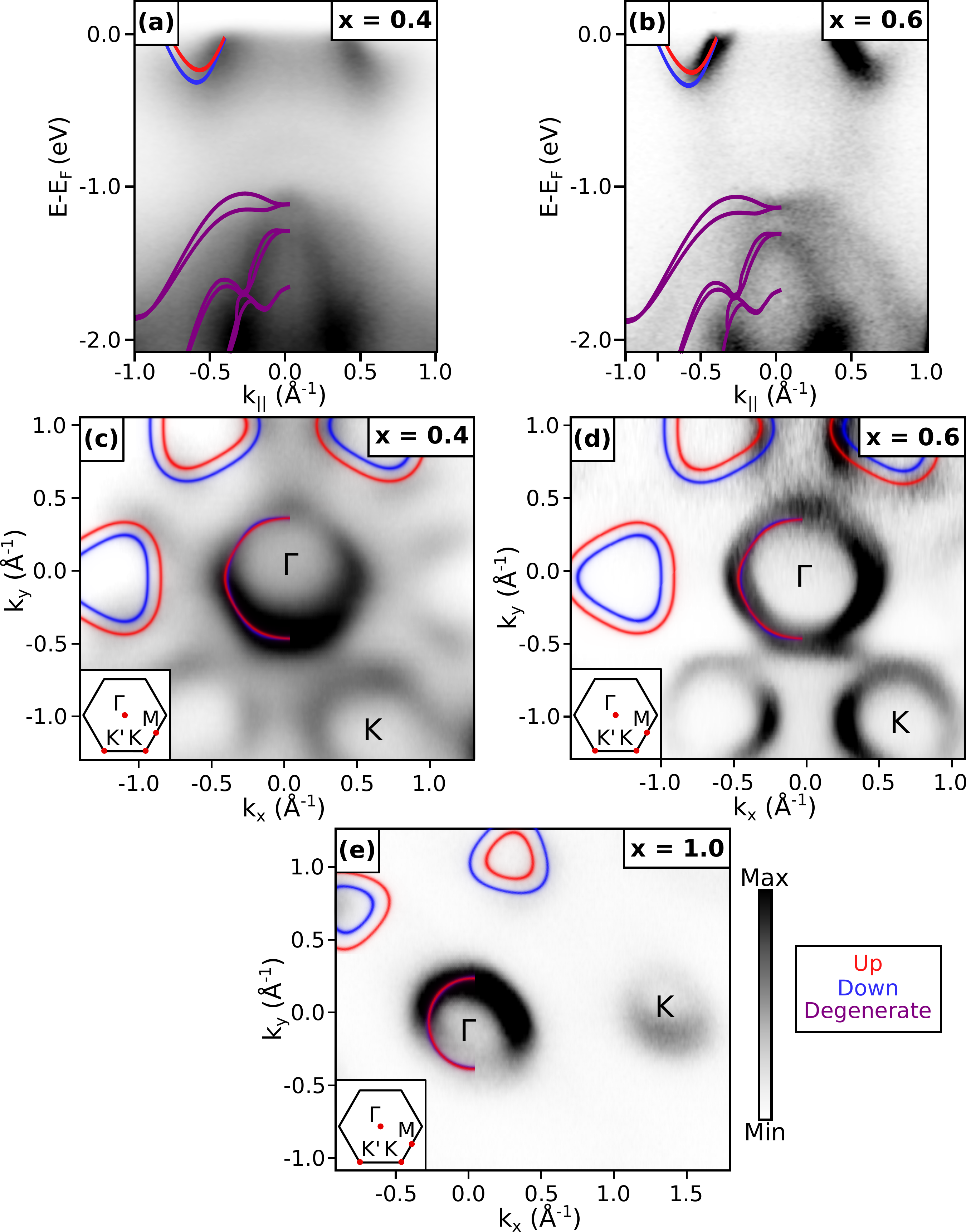} 
        \caption{a-b) ARPES spectra of $(\text{La}_{x}\text{Pb}_{1-x}\text{Se})_{1.14}(\text{NbSe}_2)_2$ for  $x = 0.4$ (a) and $x = 0.6$ (b) recorded along $K-\Gamma-K'$ high-symmetry direction. The band structure of doped monolayer NbSe$_2$ is superimposed as a guide to the eye. c–e) ARPES Fermi surface maps of $(\text{La}_{x}\text{Pb}_{1-x}\text{Se})_{1.14}(\text{NbSe}_2)_2$ for  $x = 0.4$ (c), $x = 0.6$ (d) and $x = 1.0$ (e), acquired at $h\nu = 150$ eV. Half of the Fermi surface is superimposed on each map as a guide to the eye. All compounds exhibit the characteristic NbSe$_2$ hole pockets at the $\Gamma$ and $K$ points, with both pockets shrinking as the La content in the rocksalt layer increases.}
        \label{fig:FS}
    \end{figure*}

    \begin{figure*}
        \includegraphics[scale=0.35]{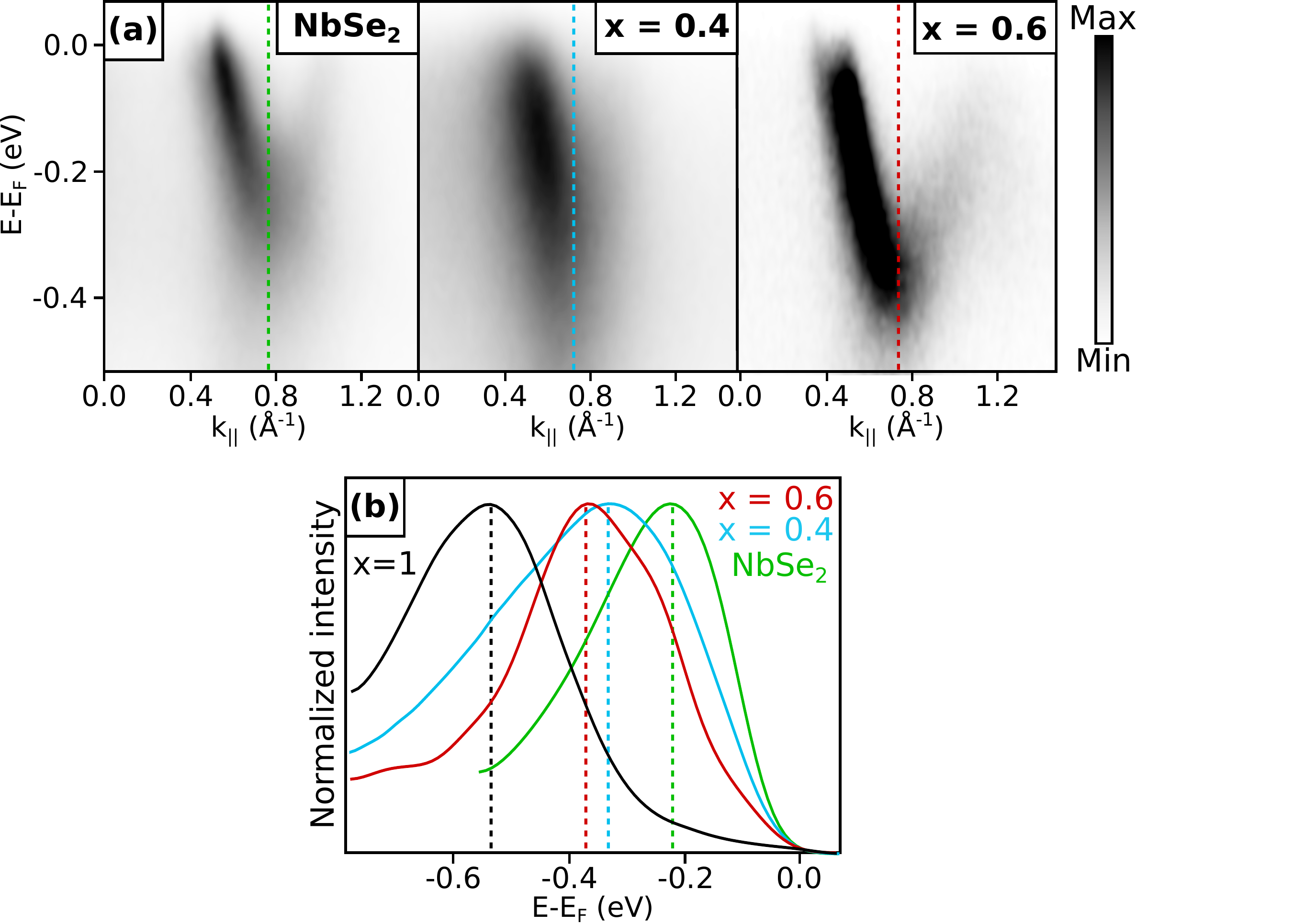} 
        \caption{a) Band dispersion of NbSe$_2$, $(\text{La}_{0.4}\text{Pb}_{0.6}\text{Se})_{1.14}(\text{NbSe}_2)_2$ and $(\text{La}_{0.6}\text{Pb}_{0.4}\text{Se})_{1.14}(\text{NbSe}_2)_2$ along the $\Gamma-K$ high-symmetry direction  b) Energy distribution curves (EDCs) extracted at the band minimum for each compound. Fermi energy shifts of $0.12$ eV (or $0.27~ \text{e}^-/$Nb) and $0.14$ eV (or $0.31~ \text{e}^-/$Nb) for $x=0.4$ and $x=0.6$, respectively, demonstrate a tunable intermediate doping regime between pristine NbSe$_2$ and the pure La-misfit limit displayed in black ($\approx 0.3$ eV) \cite{leriche_misfit_2021}.}
        \label{fig:EDC}
    \end{figure*}

 Misfit crystals are a fertile playground for the crystallography community, thanks to their rich chemistry owing to compositional flexibility and wide choices of available 2D units. 
However, guiding the chemical composition towards the construction of misfit samples with tailored physical properties is non-trivial. In our past work, we showed that this is possible by means of density functional theory calculations based on work function (W), i.e. the minimum energy required to extract an electron from a material's surface and send it to the vacuum \cite{zullo_misfit_2023}. Our results show that the rocksalts behave as electron donors thanks to their low work function. The opposite holds for the TMDs, thus, there is a preferential direction for electrons transfer in the misfit, from the rocksalts to the TMDs. 

Furthermore, it is possible to tune the charging of the TMD by appropriately choosing the rocksalt counterpart. For example, explicit calculations for several misfit surfaces terminated by a single layer of NbSe$_2$, having different RS units with comparable mismatching ratios of $7/4$, show the charge
transfer decreases by progressively decreasing the difference W(NbSe$_2$) $-$ W(RS). 

Two extreme cases are represented by the compounds with LaSe and PbSe. These two rocksalts posses respectively the highest and lowest work function among the selenides \cite{zullo_misfit_2023}. Therefore, NbSe$_2$ electronic structure is heavily doped in the La rich compound $(\text{La}\text{Se})_{1.14}(\text{NbSe}_2)_2$. Instead, the Pb rich compound $(\text{Pb}\text{Se})_{1.14}(\text{NbSe}_2)_2$ results in zero charge transfer. The two limit cases, high doping and zero doping, are met by these two compounds, as predicted from the band alignment plot.
Therefore, by partially substituting the La with Pb atoms in the rocksalt, it is feasible to engineer a misfit  surface so that the doping of the TMD is rigidly tunable, as predicted by theoretical calculations in $\text{TiSe}_2$ \cite{zullo_misfit_2023}.

Thus, we wonder if it is possible to tune rigidly the doping of $\text{NbSe}_2$ by La-Pb alloying in misfit surfaces. 
For this reason we consider MLCs having the following stoichiometry $(\text{La}_x\text{Pb}_{1-x}\text{Se})_{1.14}(\text{NbSe}_2)_2$ as a function of $x$ and lattice mismatch ratio $7/4$.
From the previous reasoning, we expect that the La concentration ($x$) allows tuning of the carrier concentration in the $\text{NbSe}_2$ layers within the misfits, with $x = 1$ ($x = 0$) corresponding to the highest (lowest) n-doping.
In Fig. \ref{fig:band}, we show the calculated band structure of the $(\text{La}_x\text{Pb}_{1-x}\text{Se})_{1.14}(\text{NbSe}_2)_2$ misfit as a function of $x$ (see Methods). In each panel, the band structure of the misfit is unfolded onto the hexagonal Brillouin Zone of the TMD. This procedure facilitates a comparison between the misfit band structures (black colormap) and the superimposed band structure of an isolated single layer $\text{NbSe}_2$ (red bands).
For all the concentrations, we report the calculated amount of charge transferred from the rocksalt to $\text{NbSe}_2$. This is achieved by integrating the density of states in a single layer of NbSe$_2$ over the energy range associated with the shift in Fermi energy caused by the misfit.

By inspecting our calculations in Fig. \ref{fig:band}, we can immediately state that the doping is increased by solely increasing $x$. 
This statement is enforced by the evolution of the electrons per Nb atoms as a function of the concentration of the La atoms.
The Fermi level of the TMD in the misfit (blue dashed line) is shifted with respect to the single layer one (red dashed line) by an amount that is directly related to the presence of La in the rocksalt.
The direction of the arrow indicates a crescent doping with increasing of the La content.

Remarkably, the bands of $\text{NbSe}_2$ inside the misfit are essentially identical to those of an isolated single layer $\text{NbSe}_2$.
This result highlights the two-dimensional nature of the TMD within the misfit heterostructure. 
From the first to the final panel of Fig. \ref{fig:band}, an increase in Pb concentration is responsible for the development of some extra RS states below the Fermi level due to the change of band alignment with NbSe$_2$.
A rigid upward shift is observed when going from the misfit with 100$\%$ of Pb atoms ($x = 0$) to the one with 100$\%$ of La atoms ($x = 1$). 
Following the direction of the arrow in Fig. \ref{fig:band}, the doping is increased rigidly much like in a field effect-transistor configuration. 
This is a crucial result, enforcing the general picture that the doping of the TMD in misfit layer compounds is tunable by chemical insight.

To experimentally prove the predicted tunability of the electronic properties, we perform angle-resolved photoemission spectroscopy (ARPES) on two distinct samples of $(\text{La}_{x}\text{Pb}_{1-x}\text{Se})_{1.14}(\text{NbSe}_2)_2$ with lanthanum concentration of $x=0.4$ and $x=0.6$ (see Methods). ARPES spectra along $K-\Gamma-K'$ for both compound are presented Fig. \ref{fig:FS}a and b. Well-defined bands originating from the Nd 4d orbitals are visible within 0.4 eV below the Fermi level ($E_F$). Additionally, bands originating from the Se 4p orbitals exhibit a dispersion maximum around -1 eV relative to $E_F$ \cite{nakata_anisotropic_2018}. We also map the Fermi surfaces (FS) of both compounds and as shown in Fig. \ref{fig:FS}c and d, both samples exhibit the characteristic FS shape of NbSe$_2$, with well-defined hole pockets centered at the $\Gamma$ and $K$ points of the Brillouin zone \cite{kundu_charge_2024}. The $x=1$ sample is also shown for comparison \cite{leriche_misfit_2021}. in this case, the pockets at $\Gamma$ and $K$ are much smaller, which directly indicates an ultra-doped regime since electron doping shifts the Fermi level upward, filling and shrinking these hole pockets. This shows that the doping is highest when the rocksalt layer contains only lanthanum, as expected from calculations.

\begin{figure*}
    \includegraphics[scale=0.35]{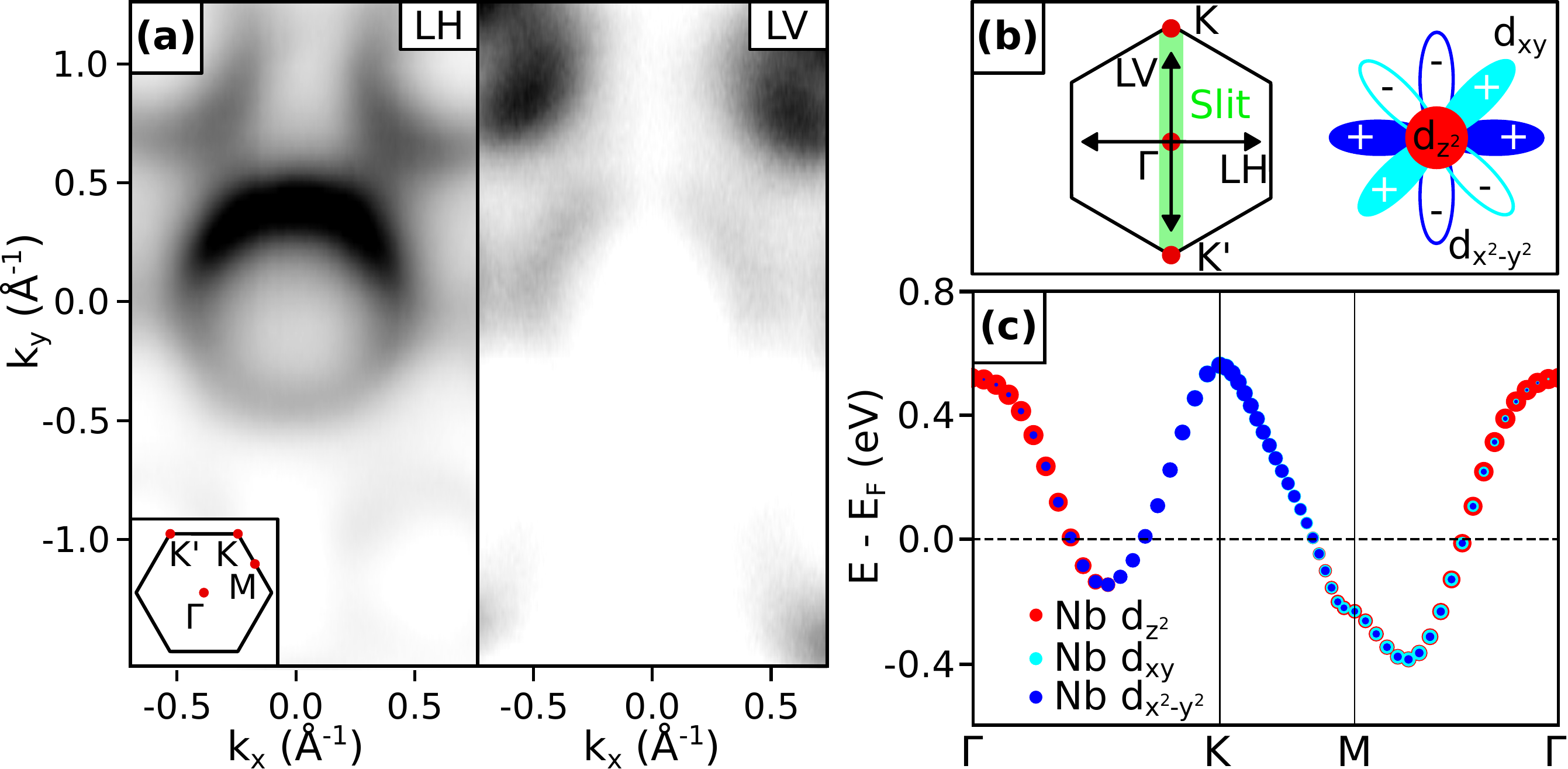} 
    \caption{a) Polarization-dependent ARPES Fermi surface of $(\text{La}_{0.4}\text{Pb}_{0.6}\text{Se})_{1.14}(\text{NbSe}_2)_2$. Under linear horizontal (LH) polarization, the $\Gamma$ pocket exhibits maximum intensity, dominated by the even-parity Nb $d_{z^2}$ orbital. In linear vertical (LV) polarization, the $\Gamma$ pocket is near-completely suppressed while the $K$ pockets remain intense due to their mixed orbital character (due to $d_{xy}$ contribution). b) Schematic of the experimental geometry and orbital symmetries. The detector slit is aligned with the $\Gamma-K$ direction of the hexagonal Brillouin zone and the arrows indicate the directions of the Linear Horizontal (LH) and Linear Vertical (LV) polarizations. On the right, the symmetries of the Nb $d_{z^2}$, $d_{xy}$, and $d_{x^2-y^2}$ orbitals are shown. c) Calculated band dispersion of 1H-NbSe$_2$ along high-symmetry paths. The colored markers represent the projected orbital character of the Nb $4d$ states, illustrating the orbital mixing of Nb 4d bands: $d_{z^2}$ and $d_{x^2-y^2}$ for $\Gamma$ pocket and $d_{xy}$ and $d_{x^2-y^2}$ for K pockets}
    \label{fig:pola}
\end{figure*}

In line with theoretical expectations that charge transfer increases with the lanthanum content in the rocksalt (RS), a precise measurement of the Nb $4$d band shift is essential to quantify the effective doping of NbSe$_2$ in these heterostructures. While the extreme doping of the pure La misfit ($x=1$) is visually obvious as presented in Fig. \ref{fig:FS}e, the analysis of intermediate compositions is more subtle. At first glance, the $\Gamma$ pockets for $x=0.4$ and $x=0.6$ appear similar in size due to spectral broadening. However, we can overcome this limit by extracting Energy Distribution Curves (EDCs) along $\Gamma - K$, to determine the Nb 4d band shift more precisely, as presented in Fig. \ref{fig:EDC}b. This analysis unveils a rigid energy shift relative to pristine NbSe$_2$. Specifically, the $x=0.4$ and $x=0.6$ compounds exhibit Fermi energy shifts of $0.12$~eV (or $0.27~ \text{e}^-/$Nb) and $0.14$~eV (or $0.31~ \text{e}^-/$Nb), respectively. As presented in Fig. \ref{fig:EDC}b, these values represent an intermediate doping regime, located between pristine NbSe$_2$ and the $0.3$~eV shift previously reported for the pure lanthanum misfit compound $(\text{LaSe})_{1.14}(\text{NbSe}_2)_2$ \cite{leriche_misfit_2021}. This successfully demonstrates that stoichiometric control of the RS layer provides a versatile method for achieving doping that is otherwise inaccessible via conventional gating techniques.

To elucidate the orbital character of the bands, we conduct polarization-dependent ARPES measurements on $(\text{La}_{0.4}\text{Pb}_{0.6}\text{Se})_{1.14}(\text{NbSe}_2)_2$, exploiting the transition matrix element selection rules. As shown in Fig. \ref{fig:pola}a, switching the polarization of the incident photon beam between linear horizontal (LH, even operator with respect to the horizontal symmetry plane) and linear vertical (LV, odd operator) polarizations reveals a clear contrast in the spectral weight. In the horizontal symmetry plane, the photoemission intensity is non-zero only if the excited initial state has the same parity as the light polarization with respect to the incident plane (horizontal), as schematically illustrated in Fig. \ref{fig:pola}b along with the corresponding orbital symmetries. Under LH even polarization, the Fermi pocket at the $\Gamma$ point (primarily derived from the even-parity Nb $d_{z^2}$ orbital) therefore exhibits maximum intensity. It is nearly completely suppressed in the LV configuration. Since the selection rule only applies in the horizontal plane, residual traces of its spectral weight is visible away from the detector slit center, aligned along $k_x$ in these measurements. Conversely, the detection of signal at the $K$ points in both channels is consistent with a mixed orbital character (including $d_{xy}$ and $d_{x^2-y^2}$ states). This signature is identical to the expected behavior of 1H-NbSe$_2$, as displayed in Fig. \ref{fig:pola}c, where the $\Gamma$ pocket is mainly derived from even-parity orbitals ($d_{z^2}$ and $d_{x^2-y^2}$) while the $K$ pocket originates from a mix of odd and even parity orbitals ($d_{xy}$ and $d_{x^2-y^2}$). These observations provide unambiguous evidence that the terminating NbSe$_2$ layer conserves its intrinsic orbital identity, remaining largely unaffected by the adjacent RS layers in these misfit heterostructures.

\begin{figure*}
    \includegraphics[scale=0.35]{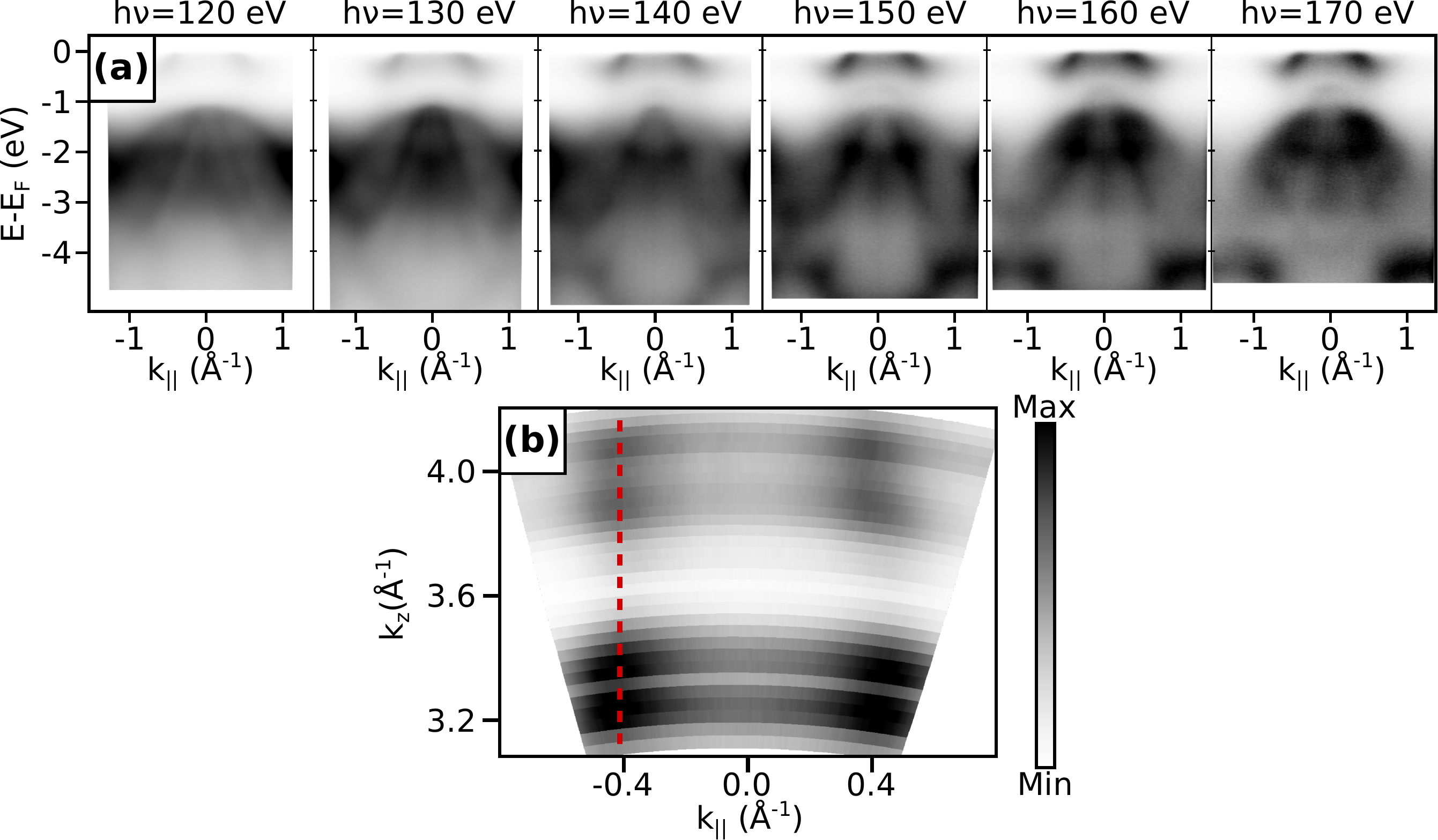} 
    \caption{a) Photon-energy-dependent ARPES spectra of $(\text{La}_{0.4}\text{Pb}_{0.6}\text{Se})_{1.14}(\text{NbSe}_2)_2$ along $\Gamma-K$ recorded between $h\nu = 120$ and $170$ eV. The Nb 4d bands maintain a constant size and shape across the entire range, whereas the valence bands show a clear $k_z$ dependence. b) Constant energy map at Fermi energy in the $(k_z, k_{//})$ plane for ARPES spectra along $\Gamma-K$ recorded  between $h\nu = 20$ and $50$ eV and calculated using $V_0=14.7~$eV. The presence of a nearly straight vertical signal with only weak modulations indicates a nearly non-dispersive electronic structure along the out-of-plane direction ($k_z$). Variations in intensity are attributed to ARPES matrix element effects.}
    \label{fig:kz}
\end{figure*}

A central question regarding misfit heterostructures is the degree to which the rocksalt (RS) layer perturbs the two-dimensional electronic properties of NbSe$_2$. Although the interface between the NbSe$_2$ and RS layers is characterized by strong iono-covalent bonding, it is essential to verify if the terminating NbSe$_2$ monolayer maintains its quasi-2D electronic identity. We address this by investigating the $k_z$ dispersion of the electronic states through photon-energy-dependent ARPES.

Measurements on $(\text{La}_{0.4}\text{Pb}_{0.6}\text{Se})_{1.14}(\text{NbSe}_2)_2$ along the $\Gamma - K$ direction for photon energies ranging from 120 to 170 eV along the $\Gamma-K$ direction are shown in Fig. \ref{fig:kz}a. Excluding that the photoemission intensity fluctuates (likely due to ARPES matrix element effects), the Nb 4d bands, located between $E_F$ and -0.4 eV, maintain a constant size and shape across the entire energy range: this is a signature of the two-dimensional character of the system. In contrast, the Se 4p bands, whose dispersion maximum is located around -1 eV relative to $E_F$, exhibit a clear $k_z$ dependence, suggesting hybridization with the RS layer. This interaction appears confined to the valence states and is unlikely to influence electronic properties, such as superconductivity or charge density waves.

To further confirm this 2D character, we conduct a systematic $k_z$ mapping at lower photon energies (20 to 50 eV). A constant-energy cut at the Fermi energy in the $(k_z, k_{//})$ plane (Fig. \ref{fig:kz}b) reveals a nearly straight vertical signal with only weak modulations. While this constant behavior along $k_z$ emphasizes a predominantly bidimensional electronic structure, the presence of these modulations reveals a small persistent 3D character. However, this out-of-plane dispersion is significantly weaker than in bulk 2H-$\text{NbSe}_2$, where a clear 3D character is well established \cite{weber_three-dimensional_2018}. The weakness of this out-of-plane dispersion provides evidence that the terminating NbSe$_2$ layer in these misfit compounds nearly retains its quasi-2D band structure despite the heavy doping and proximity to the RS lattice.

\section{Conclusions} \label{sec:Analysis}
   
The experimental realization of tunable doping in $(\text{La}_{x}\text{Pb}_{1-x}\text{Se})_{1.14}(\text{NbSe}_2)_2$ via lead substitution represents a significant opportunity in the engineering of electronic phases in two-dimensional TMDs. Our ARPES measurements provide a direct confirmation of the theoretical band structure calculations presented in this work. Specifically, the observed rigid binding energy shifts of $0.12$~eV and $0.14$~eV for $x=0.4$ and $x=0.6$, respectively, are in good agreement with our DFT calculations of the Fermi level shifts. This quantitative match validates the "field-effect transistor" model, where the rocksalt layer acts as a donor whose charge transfer is precisely governed by its stoichiometry.

The preservation of the quasi-2D character of the Nb 4d band, evidenced by the weakness of $k_z$ dispersion, further supports the theoretical picture of a decoupled TMD layer despite the strong iono-covalent bonding at the RS/TMD interface. By modulating the La/Pb ratio, we have successfully filled the experimental gap between pristine NbSe$_2$ and the heavily doped $(\text{LaSe})_{1.14}(\text{NbSe}_2)_2$. This demonstrates that the electronic structure remains robust against lead substitution, behaving as a rigidly doped $1$H monolayer as predicted.

Furthermore, the ability to reach these intermediate doping levels through stoichiometric control offers a unique experimental route to study the evolution of electronic phases. While the extreme doping in the pure La-misfit suppresses the $3 \times 3$ CDW \cite{zullo_charge_2024}, our results establish a platform to track this transition and explore the quantum phase diagram. 

In conclusion, by combining theoretical predictions and experimental validation, our work demonstrates that chemical alloying in misfit layer compounds is a reliable and powerful tool for the design of 2D TMDs, providing a level of control over carrier density that is unattainable with conventional gating techniques.

\section{Methods} \label{sec:Samples}
    \paragraph{Synthesis:}
    Single crystals of $(\text{La}_{x}\text{Pb}_{1-x}\text{Se})_{1.14}(\text{NbSe}_2)_2$ were prepared by the solid-state reaction of the elemental precursors (i.e, La, Pb, Nb, Se) and subsequent chemical vapor transport using I$_2$. Under inert atmosphere, submillimeter-sized La powder was freshly scraped from the ingot (Strem Chemicals, 99.9\%) and mixed with Pb (Sigma Aldrich, 99.9\%), Se (Alfa Aesar, 99.999\%) and Nb powder (Puratronic, 99.99\%) in molar ratio La/Pb/Nb/Se = 1.14x/1.14(1-x)/2/5.14 for x = 0.40 and 0.60. The mixture was manually ground on an agate mortar and transferred into a silica tube, which was subsequently evacuated to 10$^{-3}$ Torr and sealed by flame. The mixture was firstly heated to 200 °C at a rate of 50 °C/h and held for 12 h. Then the temperature was raised to 1000 °C at the same rate and heated for 240 h before the furnace was subject to radiative cooling. The lustrous black powder (ca., 1000 mg) obtained for each reaction was placed in the clean silica tube (length: 15 cm) together with 50 mg of iodine (Aldrich, 99.9\%), followed by the evacuation at liquid nitrogen temperature (77 K) and subsequent flame sealing. The reaction mixture was loaded into the furnace designed to create temperature gradient: the lump of the mixture (x = 0.40 and 0.60) was heated at 900 °C on one side of the tube and the other side was held around 750 °C. After 240 h of thermal treatment, the reaction mixture was removed from the furnace. The single crystals grown on the wall of the tube were washed with water and ethanol and stored in vacuum. The compositional integrity of the obtained crystals was checked by their backscattered electron images and energy-dispersive X-ray (EDX) spectra acquired on scanning electron microscope (JEOL JSM IT510).
    \\
    \\
    \paragraph{Experiment:}
    ARPES measurements were performed at the CASSIOPEE beamline of the Synchrotron SOLEIL (France) using a Scienta R4000 hemispherical analyzer. All the data were acquired at a sample temperature of $17$~K. Samples were cleaved $\textit{in situ}$ under ultra-high vacuum ($P<5\times 10^{-11}$ mbar) at 17 K . The preferential cleavage between NbSe$_2$ layers is driven by the weakness of the van der Waals forces compared to the strong ionocovalent bonding at the interface between the RS and TMD layers, ensuring a clean pristine terminating NbSe$_2$ layer. The local chemical composition is systematically verified by X-ray photoelectron spectroscopy (XPS) immediately before the ARPES measurements to confirm the presence of the rocksalt layer beneath the surface, ensuring that the probed areas are not bulk NbSe$_2$ since intergrowth can occur between misfit planes during the synthesis.
    \\
    \\
    \paragraph{Theory:} 
    In \textit{ab initio} calculations, misfit surfaces can be modeled as slabs: an alternate stacking sequence of one rocksalt bilayer sandwiched between two TMD monolayers along the c-axis (TMD-RS-TMD stacking sequence). This sequence assures the neutrality of the cell, with the same charge transfer of the bulk.
    
    In the following, the TMD (RS) will be labeled as the subsystem $1$ ($2$). To construct misfit surfaces, we first optimize each subsystems separately. Then, we assemble the misfit by stacking the subsystems in the TMD-RS-TMD sequence. Finally we perform geometrical optimization of the misfit slab.
    The lattice parameter mismatch in one of the in-plane directions makes the misfit cell incommensurate. It is possible to simulate an approximate commensurate cell by considering the ratio between the subsystem's lattice parameter along the misfit axis |$\boldsymbol{a_{2}}$|/|$\boldsymbol{a_{1}}$| = 6/3.437 ( $\approx$ 7/4), and thus $\boldsymbol{a}$ =7|$\boldsymbol{a_{1}}$| $\approx$ 4|$\boldsymbol{a_{2}}$|.
    The two subunit cells in the $7\times4$ periodic approximant of $(\text{La}\text{Se})_{1.14}(\text{NbSe}_2)_2$ ($(\text{Pb}\text{Se})_{1.14}(\text{NbSe}_2)_2$) are listed below. The $\text{NbSe}_2$ sublattice has an orthorhombic cell with in-plane lattice vectors $\boldsymbol{a_{1}}$ = 3.437 \AA \ and $\boldsymbol{b_{1}}$ = 6 \AA \, while the LaSe (PbSe) sublattice has an orthorhombic cell with in-plane lattice vectors $\boldsymbol{a_{2}}$ = 6.019 (6.028) \AA \ and $\boldsymbol{b_{2}}$ = 6 \AA \ .
    
    The resulting misfit crystal has an orthorhombic cell with lattice parameters $\boldsymbol{a}$ = 24 \AA \ , $\boldsymbol{b}$ = 6 \AA. The structure has a P1 symmetry and includes $116$ atoms in the slab. Starting from these parameters, we construct the $(\text{La}_{x}\text{Pb}_{1-x}\text{Se})_{1.14}(\text{NbSe}_2)_2$ misfit series by partially substituting Pb atoms, as established in Ref. \cite{zullo_misfit_2023}. Therefore, $x$ ($1-x$) corresponds to the concentration of La (Pb) atoms in the slab.
    
    We calculate the electronic properties of the $(\text{La}_{x}\text{Pb}_{1-x}\text{Se})_{1.14}(\text{NbSe}_2)_2$ misfit series by means of density functional theory (DFT) as implemented in the quantum ESPRESSO (QE) code \cite{QE}. We employ a $4\times1\times1$ Monkhorst-Pack k-points grid and a Gaussian smearing of $0.025$ Ry for Brillouin Zone (BZ) sampling. We use the generalized gradient approximation in the Perdew–Burke–Ernzerhof \cite{PBE} parametrization for the exchange-correlation functional. In the calculation of surfaces, as the interaction among transition metal dichalcogenides layers is missing and only covalent bonds among the rocksalt and the transition metal dichalcogenides are present, we did not consider any Van der Waals correction. We consider the pseudopotential configurations taken from the Vanderbilt \cite{VANDERBILT} and PSlibrary \cite{PSLIBRARY} distributions. The values of kinetic energy cutoff for plane-wave basis set were set to $40$ Ry and for charge density $480$ Ry respectively for all misfit compounds. The atomic positions of the slab are relaxed, by means of the Broyden-Fletcher-Goldfarb-Shanno (BFGS) algorithm, with a convergence threshold of $10^{-4}$ Ry on the total energy difference between consecutive structural optimization steps and of $10^{-3}$ Ry/Bohr on all force components. All the surfaces comprise $\sim 24$\ \AA \ vacuum space to prevent interactions between periodic replicas. To capture the band-splittings of NbSe$_2$, spin-orbit coupling (SOC) was included in all the electronic structure calculations.

\bibliographystyle{apsrev4-1}
\bibliography{references}

\end{document}